\documentclass{mn2e}
\usepackage{graphicx}

\title[Transient growth and coupling of vortex and wave modes in self-gravitating gaseous discs]
{Transient growth and coupling of vortex and wave modes in
self-gravitating gaseous discs}
\author[G. R. Mamatsashvili and G. D. Chagelishvili] {G. R. Mamatsashvili $^{1}$\thanks{E-mail:
g.mamatsashvili@astro-ge.org} and G. D. Chagelishvili
$^{1}$\\
$^{1}$ E. Kharadze Georgian National Astrophysical Observatory, 2a
Kazbegi Ave., Tbilisi 0160, Georgia}
\begin{document}

\date{Accepted 2007 July 26. Received 2007 July 18; in original form 2007 January 28}

\pagerange{\pageref{firstpage}--\pageref{lastpage}} \pubyear{2007}

\maketitle

\label{firstpage}

\begin{abstract}
Flow nonnormality induced linear transient phenomena in thin
self-gravitating astrophysical discs are studied in the shearing
sheet approximation. The considered system includes two modes of
perturbations: vortex and (spiral density) wave. It is shown that
self-gravity considerably alters the vortex mode dynamics -- its
transient (swing) growth may be several orders of magnitude stronger
than in the non-self-gravitating case and 2-3 times larger than the
transient growth of the wave mode. Based on this finding, we comment
on the role of vortex mode perturbations in a gravitoturbulent
state. Also described is the linear coupling of the perturbation
modes, caused by the differential character of disc rotation. The
coupling is asymmetric -- vortex mode perturbations are able to
excite wave mode ones, but not vice versa. This asymmetric coupling
lends additional significance to the vortex mode as a participant in
spiral density waves and shocks manifestations in astrophysical
discs.
\end{abstract}


\section{Introduction}

The comprehension of the regular spiral structure of galaxies
provided a powerful incentive for the investigation of dynamical
processes and phenomena in astrophysical discs. Afterwards the study
in this direction intensified and somewhat changed (emphases
shifted) -- in the 70s of the last century a powerful branch of
astrophysical disc research was developed along with the X-ray
astronomy. The latter revealed a strong energy release in discs --
accretion of the disc matter onto its centre -- that is ascribed to
anomalous viscosity due to some sort of turbulence \cite{SS73,Pr81}.
In self-gravitating discs, turbulence and angular momentum transport
process are commonly attributed to spiral density (SD) waves
\cite{G01,LR04,LR05,M05}. In other words, in the study of
self-gravitating discs, attention is focused on the dynamical
activity (possibility of amplification) of SD wave perturbations. On
the other hand, in investigating turbulence and angular momentum
transport in non-self-gravitating discs, the main emphasis is put on
the dynamical activity of vortical perturbations
\cite{LCC88,GL99,Lo99,DSC00,IK01,Ta01,Da02,Ch03,KB03,Te03,UR04,Ye04,AMN05,Bo05,JG05}.
Interest to this type of perturbations is also connected with the
idea that long-lived vortices in protoplanetary discs play an
important role in the planet formation
\cite{BS95,Br99,GL99,GL00,KB06}. Regular vortex structures are also
observed in several spiral galaxies \cite{FK99}.

In general, one can say that the astrophysical disc community has
been advancing along a crooked path solving dynamical problems.
Matters are complicated by one of the main sources supplying
energetically dynamical processes in discs -- differential character
of disc rotation. Moreover, this source is universal, as
astrophysical discs actually always rotate differentially. The
complication is due to the nonnormal character of flows with
inhomogeneous kinematics (such as differential rotation). The
nonnormality and its consequences were well understood and precisely
described by the hydrodynamic community in the 90s of the last
century. The imperfection of the traditional/modal analysis
(spectral expansion of perturbations in time and subsequent
examination of eigenfunctions) in regard to smooth (without
inflection point) inhomogeneous/shear flows was revealed: operators
existing in the mathematical formalism of the modal analysis of
shear flows (e.g. plane Couette and Poiseuille) are nonnormal and,
hence, corresponding eigenfunctions are nonorthogonal and strongly
interfere \cite{RSH93,Tr93}. Consequently, a correct approach should
fully analyse the interference of eigenfunctions, which is actually
calculable/manageable for asymptotically large times. In fact, in
the modal analysis the main focus is on the asymptotic stability of
flows, while no attention is directed to any particular initial
value or finite time period of the dynamics -- this period of the
evolution is thought to be of no significance and is left to
speculation. This fact prompted the above mentioned revision of the
generally accepted spectral/modal approach with the special emphasis
being shifted from the analysis of long time asymptotic flow
stability to the study of finite time (transient) behaviour. It was
demonstrated that just because of this nonnormality of the
operators, a strong linear transient growth of vortex and/or wave
mode perturbations occurs in asymptotically stable (in accordance
with Rayleigh's stability criterion; Rayleigh 1880) hydrodynamic
shear flows \cite{Gu91,BF92,RH93,FI93,FI00}, that determines the
fate of a flow.

Differential rotation (i.e., the disc flow nonnormality) is
determinant in a number of cases. Even when other basic factors
(e.g. self-gravity, stratification) are involved, it interplays with
them and strongly modifies the dynamical picture. This circumstance
attaches great importance to disc flow shear (nonnormality) induced
phenomena. {\it The main goal of this paper is to study the
dynamical manifestations of the nonnormality in a simple model of
self-gravitating astrophysical discs.} Specifically, we consider the
linear dynamics of two-dimensional perturbations in thin
self-gravitating gaseous discs in the shearing sheet approximation
\cite{GLB65}. This system includes two modes of perturbations --
vortex and (spiral density) wave. The distinguishing features of
these modes are the following: the vortex mode is aperiodic and has
nonzero potential vorticity; the SD wave mode is oscillatory and has
zero potential vorticity. We use the nonmodal approach instead of
modal/spectral one in describing the linear dynamics of
perturbations. This approach was applied in previous well-known
papers \cite{GLB65,JT66,GT78,To81}, but they concentrate on
perturbations with zero potential vorticity, i.e., on SD wave mode
perturbations, whereas vortex mode ones can be important as well,
for example, in the angular momentum transport. This is evidenced by
the references above concerning the dynamical activity of vortical
perturbations in astrophysical, though non-self-gravitating, discs.
We would like to emphasize that associating turbulence and angular
momentum transport in self-gravitating discs only with SD waves, one
in that way does not make rigorous identification (according to the
value of potential vorticity) of perturbation modes and does not
investigate the relative contributions/fractions of vortical and SD
wave perturbations in shear stresses. In other words, the role of
vortical perturbations is left out of self-gravitating disc analysis
(see also Sec. 5). Considered in the present paper linear coupling
of SD wave and vortex modes (see below) makes the latter more
obvious participant in the angular momentum transport process, at
least as an additional generator of SD waves.

We outline some possible ways of injection of nonzero vorticity
perturbations discussed in the literature. Spatial inhomogeneities
of entropy (temperature) distribution, that is, baroclinic
instability \cite{KB03,K04} and Rossby wave instability \cite{Lo99},
are able to generate nonzero vorticity. Random perturbations of the
vorticity field may be present in a disc that forms as a result of
collapse of a turbulent protostellar cloud. Accretion of clumps of
gas onto a disc and convection are other possibilities for vorticity
generation (see Godon \& Livio 2000 for the details about the last
three mechanisms).

The first dynamical manifestation of the disc flow nonnormality is a
transient character of growth of both vortex and wave mode
perturbations irrespective of the value of Toomre's stability
parameter $Q$. In dynamically important regions of wavenumber plane,
\emph{vortex mode perturbations always exhibit larger growth factors
than wave mode ones.} Consequently, the underestimation of the
vortex mode and its transient (swing) growth may result in an
incomplete dynamical picture of discs. First of all, we would like
to mention the following from our investigation fact: the transient
growth of vortex mode perturbations in self-gravitating discs is
much stronger than in non-self-gravitating ones. Due to this fact,
the presence of self-gravity (gravitational instability) might, in
principle, allow for the onset of hydrodynamic turbulence in
astrophysical discs. The so-called \emph{bypass} mechanism of the
onset of hydrodynamic turbulence, which was elaborated by the
hydrodynamic community in the 90s \cite{GG94,HR94,BDT95,Gr00}, may
play a part in the process of triggering turbulence in
self-gravitating discs as well, because linear transient
amplification of perturbations due to flow nonnormality, which can
supply turbulence with mean flow energy, is a basis for this
concept. The details of the bypass concept as applied to
astrophysical discs are given in Chagelishvili et al. 2003 and in
Tevzadze et al. 2003.

The second implication of flow nonnormality -- the linear coupling
of vortex and wave modes, which is caused by the strong shear of
disc flow -- is also important for the proper comprehension of disc
flow dynamics. One should note that the coupling is asymmetric:
vortex mode perturbations (i.e. perturbations with nonzero potential
vorticity) are able to excite SD waves (i.e. perturbations with zero
potential vorticity), but not vice versa. So, this asymmetric
coupling lends additional significance to the vortex mode as a
participant in SD waves and shocks manifestations in astrophysical
discs \cite{Bo05,Bo07}.

The paper is organized as follows: physical approximations and the
mathematical formalism of the problem are introduced in Sec. 2,
classification of perturbations is described in Sec. 3, the
numerical analysis of the linear dynamical equations, including
transient growth and coupling of vortex and wave modes, is presented
in Sec. 4, discussions and summary are given in Sec. 5.

\section{Physical Model and Equations}

Let us study the linear dynamics of vortex and wave mode
perturbations in a simple analogue to a differentially rotating disc
-- in a thin self-gravitating gas sheet, where unperturbed velocity
field is a parallel flow with a constant shear (the shearing sheet
approximation). A Coriolis force is included to take into account
the effects of disc rotation. The equation of state is assumed to be
polytropic. In this case, the linearized dynamical equations read as
\cite{GT78}:
\begin{equation}
\frac{\partial \sigma}{\partial t} + 2Ax \frac{\partial
\sigma}{\partial y} + {\Sigma_0} \left(\frac {\partial u}{
\partial x} + \frac{\partial v}{\partial y} \right) = 0,
\end{equation}
\begin{equation}
\frac{\partial u}{\partial t} + 2Ax \frac{\partial u}{\partial y} -
2\Omega_0 v =- \frac{\partial}{\partial x} \left(
c_{s}^2\frac{\sigma}{\Sigma_0}+\psi \right)
\end{equation}
\begin{equation}
\frac{\partial v}{\partial t} + 2Ax \frac{\partial v}{\partial y} +
2Bu=-\frac{\partial}{\partial y} \left(
c_{s}^2\frac{\sigma}{\Sigma_0}+\psi \right).
\end{equation}
This set of linear perturbation equations is supplemented by
Poisson's equation
\begin{equation}
\Delta \psi=4\pi G \sigma\delta(z).
\end{equation}
Here $u,v,\sigma$ and $\psi$ are, respectively, the perturbed radial
and azimuthal velocities, surface density and gravitational
potential of the gas sheet with spatially constant unperturbed
surface density $\Sigma_0$. $\Omega_0$ is the angular velocity of
the shearing sheet, $c_{\rm s}$ is the sound speed in the gas, $x$
and $y$ are, respectively, the radial and azimuthal coordinates of
the shearing sheet, $A$ is the Oort constant (shear parameter),
which is $A/\Omega_0 \simeq -0.75 < 0$ for quasi-Keplerian rotation
considered in this paper, and $B\equiv A+\Omega_0$. As usual, we
define the epicyclic frequency $\kappa$ by $\kappa^2\equiv
4B\Omega_0.$

Following the standard procedure of nonmodal analysis
\cite{GLB65,GT78,NS92,Ch97}, we introduce the spatial Fourier
harmonics (SFHs) of perturbations with time-dependent phases:
\begin{equation}
F({\bf r},t)\sim F(k_x,k_y,t) \exp [ {\rm i}k_x(t)x + {\rm i}k_yy],
\end{equation}
\[
k_x(t) = k_x - 2A k_y t,
\]
where $F\equiv(u,v,\sigma,\psi)$. The streamwise/azimuthal
wavenumber $k_{y}$ remains unchanged. The streamcross/radial
wavenumber $k_{x}(t)$ changes with time at a constant rate due to
the effect of the shearing background on wave crests. One can say
that in the linear approximation SFHs ``drift'' along the
$k_{x}$-axis in ${\bf k}$-plane (wavenumber plane). In other words,
lines of constant phase of each SFH are sheared over by the basic
flow in physical plane. So, imposing initially any kind (vortex
or/and wave mode) of a leading SFH ($k_{x}(0)/k_{y}<0$) on the flow,
in the linear regime, it eventually becomes a trailing one
($k_{x}(t)/k_{y}>0$) as time passes. It should also be noted that
SFHs represent the simplest/basic ``elements'' of dynamical
processes at constant shear rate and greatly help to grasp the
phenomena of transient growth and coupling of perturbation modes.

Substituting equation (5) into equations (1-4) and introducing
$\hat{\sigma} \equiv i\sigma/\Sigma_0,~ \phi \equiv i\psi$, we get
the system of ordinary differential equations that govern the linear
dynamics of SFHs of perturbations:
\begin{equation}
{\frac {d {\hat {\sigma}}(t)} {d t}} = k_x(t) u(t) + k_y v(t).
\label{eqsigma2}
\end{equation}
\begin{equation}
{\frac{d u(t)} {d t}} - 2\Omega_0 v(t)= - k_x(t) \left[c_{\rm s}^2
{\hat {\sigma}}(t) + \phi(t)\right], \label{equ2}
\end{equation}
\begin{equation}
{\frac {d v(t)}{d t}} +2Bu(t) =- k_y \left[c_{\rm s}^2 {\hat
{\sigma}}(t) + \phi(t)\right], \label{eqv2}
\end{equation}
\begin{equation}
\phi(t)=-\frac{2\pi G\Sigma_0}{k(t)} {\hat {\sigma}}(t),
\label{eqvphi}
\end{equation}
where $k(t)=(k_x^2(t)+k_y^2)^{1/2}$. The last equation follows from
Poisson's equation and is straightforward to derive
\cite{GT78,NS92}.

One can easily show that this system possesses an important time
invariant:
\begin{equation}
k_x(t) v(t) - k_y u(t) + 2B{\hat {\sigma}}(t) \equiv {\cal I,}
\end{equation}
which follows (for SFHs of perturbations) from the conservation of
potential vorticity. This time invariant ${\cal I}$, in turn,
indicates the existence of the vortex/aperiodic mode in the
perturbation spectrum. Clarification of the role of this mode in the
disc flow dynamics represents the primary purpose of our study.

In the calculations below, we use the quadratic form (spectral
energy density) for a separate SFH as a measure of its intensity:
\begin{equation}
E(t)\equiv \frac{\Sigma_0}{2} \left (|u|^2+|v|^2
\right)+\frac{\Sigma_0}{2} c_s^2 {|\hat {\sigma}|}^2,
\end{equation}
where the two terms correspond to the kinetic and potential energies
of SFH, respectively. Strictly speaking, this is not an exact
expression for perturbation energy, since it does not contain terms
corresponding to gravitational energy. Nevertheless, we find this
quadratic form convenient for a comparative analysis of transient
growth of perturbation modes at different values of Toomre's
parameter $Q$ presented below.

The numerical study of SFH's dynamics is based on equations (6-11).
However, for the comprehension of the dynamical processes --
transient growth and coupling of vortex and SD wave modes and -- it
is advisable to rewrite them in the form of a single second order
inhomogeneous differential equation for $\phi$. Introducing
dimensionless parameters and variables: $ \tau \equiv t\kappa,~~ K_y
\equiv c_{\rm s}k_{y}/\kappa,~~K_x(\tau) \equiv c_{\rm
s}k_x(t)/\kappa,~~K(\tau) \equiv c_{\rm
s}k(t)/\kappa,~~\hat{\Omega}_{0}\equiv \Omega_{0}/\kappa\simeq
1,~~\hat{A} \equiv A/\kappa\simeq -0.75,~~\hat{B}\equiv
B/\kappa\simeq 0.25,~~\hat{\cal I} \equiv {\cal I}/\kappa,~~Q\equiv
c_{\rm s}\kappa/\pi G \Sigma_{0},~~\hat{u} \equiv u/c_{\rm
s},~~\hat{v} \equiv v/c_{\rm s},~~\hat{\phi} \equiv \phi/c^2_{\rm
s},~~\hat{E}(\tau) \equiv E(t)/\Sigma_{0}c^2_{\rm s},$ one finally
gets:
\begin{equation}
K_x(\tau) {\hat v}(\tau) - K_y {\hat u}(\tau) + 2\hat{B}{\hat
{\sigma}}(\tau) \equiv {\hat {\cal I}},
\end{equation}
\begin{equation}
{\hat E}(\tau) \equiv \frac{1}{2} \left (|{\hat u}|^2+|{\hat v}|^2
+|\hat {\sigma}|^2 \right),
\end{equation}
\begin{equation}
{\frac{d^2 {\hat{\phi}}(\tau)}{d {\tau}^2}} +
\hat{\omega}^{2}(\tau)\hat{\phi}(\tau)=-\frac{4}{QK(\tau)}
\left(\hat{\Omega}_{0}+\frac{2\hat{A}K_y^2}{K^2(\tau)}\right){\hat{\cal
I}},
\end{equation}
where
\begin{equation}
\hat {\omega}^{2}(\tau) =1+K^2(\tau)-\frac{2}{Q}K(\tau)+\frac
{12\hat{A}^{2}K_y^4}{K^4(\tau)}+\frac{8\hat{\Omega}_{0}\hat{A}K_y^2}{K^2(\tau)}.
\end{equation}
We retain $\hat{A}$ (equal to $-0.75$) in equations (14) and (15) to
make obvious the role of the flow shear in the perturbation
dynamics. The rhs of equation (14) can be viewed as a source term
for SD waves (see below) and it resembles the bar term in Goldreich
\& Tremaine 78. All other perturbed quantities are easily expressed
in terms of $\hat{\phi}(t)$ and its time derivative. We do not give
those expressions here. Notice that due to the varying radial
wavenumber $K_x(\tau)$, frequency $\hat{\omega}$ is also
dime-dependent and, as a result (we will show that below), the modes
of perturbations appear to be coupled in the linear theory.
\begin{figure}
\includegraphics[width = \columnwidth]{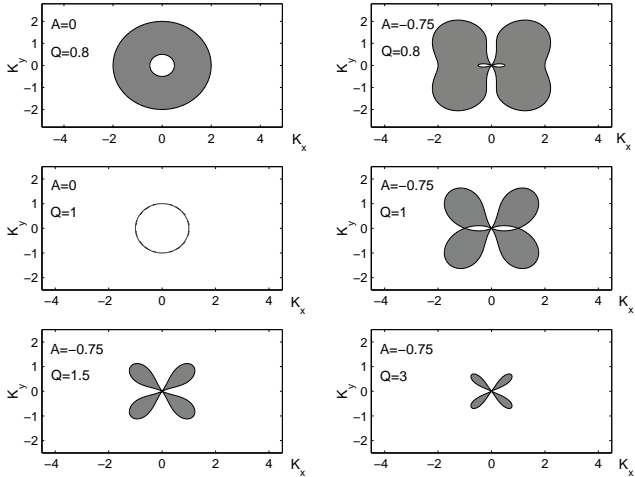}
\caption{Unstable ($\hat{\omega}^2<0$) domains (grey) in {\bf
K}-plane for various values of $Q$. At $Q<1$, unstable domains exist
even in the shearless limit. At $Q \geq 1$, compared with $A=0$
case, shear leads to the emergence of unstable domains. The last two
panels are given only for $A\neq 0$, since there are no unstable
domains in the shearless limit at $Q \geq 1$.}
\end{figure}

For further reference, in Fig.~1 we show unstable
($\hat{\omega}^2<0$) domains in {\bf K}-plane for various $Q$. At
$Q<1$, the unstable domains exist even in the shearless limit, while
at $Q \geq 1$, their occurrence is brought about just by the
combined action of shear and self-gravity.

\section{Classification of perturbations}

One can classify perturbation modes involved in equation (14) (or in
equations (6-9)) from the mathematical and physical standpoints
separately.

Mathematically, a general solution of equation (14) can be written
as a sum of two parts: a \emph{general} solution of the
corresponding homogeneous equation (oscillatory SD wave mode) and a
\emph{particular} solution of this inhomogeneous equation. It should
be emphasized that the particular solution is not uniquely
determined: the sum of a particular solution of the inhomogeneous
equation and any particular solution of the corresponding
homogeneous equation (i.e. wave mode solution) is also a particular
solution of the inhomogeneous equation, i.e., a particular solution
may comprise any amount of the wave mode.

Physically, equation (14) describes two different modes/types of
perturbations:\\ {\bf(a)} SD wave mode ($\hat{\phi}^{\rm (w)}$),
that is determined by general solution of the corresponding
homogeneous equation and has zero potential vorticity
($\hat{\cal I}=0$);\\
{\bf(b)} Vortex mode ($\hat{\phi}^{\rm (v)}$), originating from the
equation inhomogeneity (the rhs of equation (14)), is associated
with the nonoscillatory part of a particular solution of the
inhomogeneous equation. In the shearless limit, the vortex mode is
independent of time and has zero velocity divergence, but nonzero
potential vorticity. However, in the presence of a shear it acquires
divergence as well (this question is addressed in detail below).
From the above argument, it follows that the correspondence between
the aperiodic vortex mode and the particular solution of the
inhomogeneous equation is quite unambiguous -- the vortex mode is
associated only with that part of a particular solution not
containing any oscillations. The amplitude of the vortex mode is
proportional to $\hat{\cal I}$ and goes to zero when $\hat{\cal
I}=0$.
\begin{figure}
\includegraphics[width=\columnwidth]{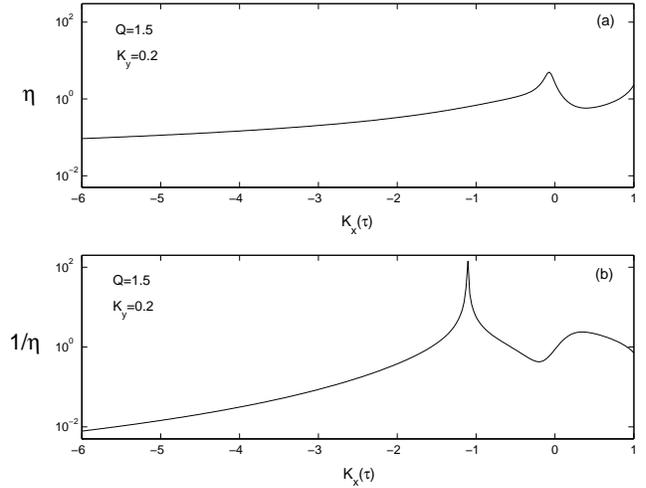}
\caption{Panel (a) is the evolution of $\eta$ in the case where a
leading pure SD wave mode SFH with only one sign (positive) of
frequency is inserted initially into equations (6-9). This wave mode
SFH acquires curl at about the time when it starts to enter the
unstable (nonadiabatic) domain in Figs.~1,4. Panel (b) shows the
evolution of $1/\eta$ for an initially inserted leading pure vortex
mode SFH. This vortex mode SFH acquires divergent nature at about
the same time. In both figures, $Q=1.5$ and $K_y=0.2$.}
\end{figure}

In the following, we will keep to the physical standpoint of
separation of perturbation modes. Thus, any solution of equations
(6-9) can be expressed as a superposition of oscillatory/SD wave and
aperiodic/vortex modes:
\[
~~~~~~~~~~~~~~~~~~~~~\hat{u}={\hat u}^{\rm (w)}+{\hat u}^{\rm (v)},~
{\hat v}={\hat v}^{\rm (w)}+{\hat v}^{\rm (v)},~\]
\[
~~~~~~~~~~~~~~~~~~~~~{\hat \sigma}={\hat \sigma}^{\rm (w)}+{\hat
\sigma}^{\rm (v)},~{\hat \phi}={\hat \phi}^{\rm (w)}+{\hat
\phi}^{\rm (v)},
\]
where ${\hat u}^{\rm (w)},~{\hat v}^{\rm (w)},~{\hat \sigma}^{\rm
(w)}$ and ${\hat u}^{\rm (v)},~{\hat v}^{\rm (v)},~{\hat
\sigma}^{\rm (v)}$ are found from ${\hat \phi}^{\rm (w)}$ and ${\hat
\phi}^{\rm (v)}$ respectively.

\begin{figure*}
\begin{minipage}[t]{\textwidth}
\includegraphics[width=0.95\textwidth, height=0.4\textwidth]{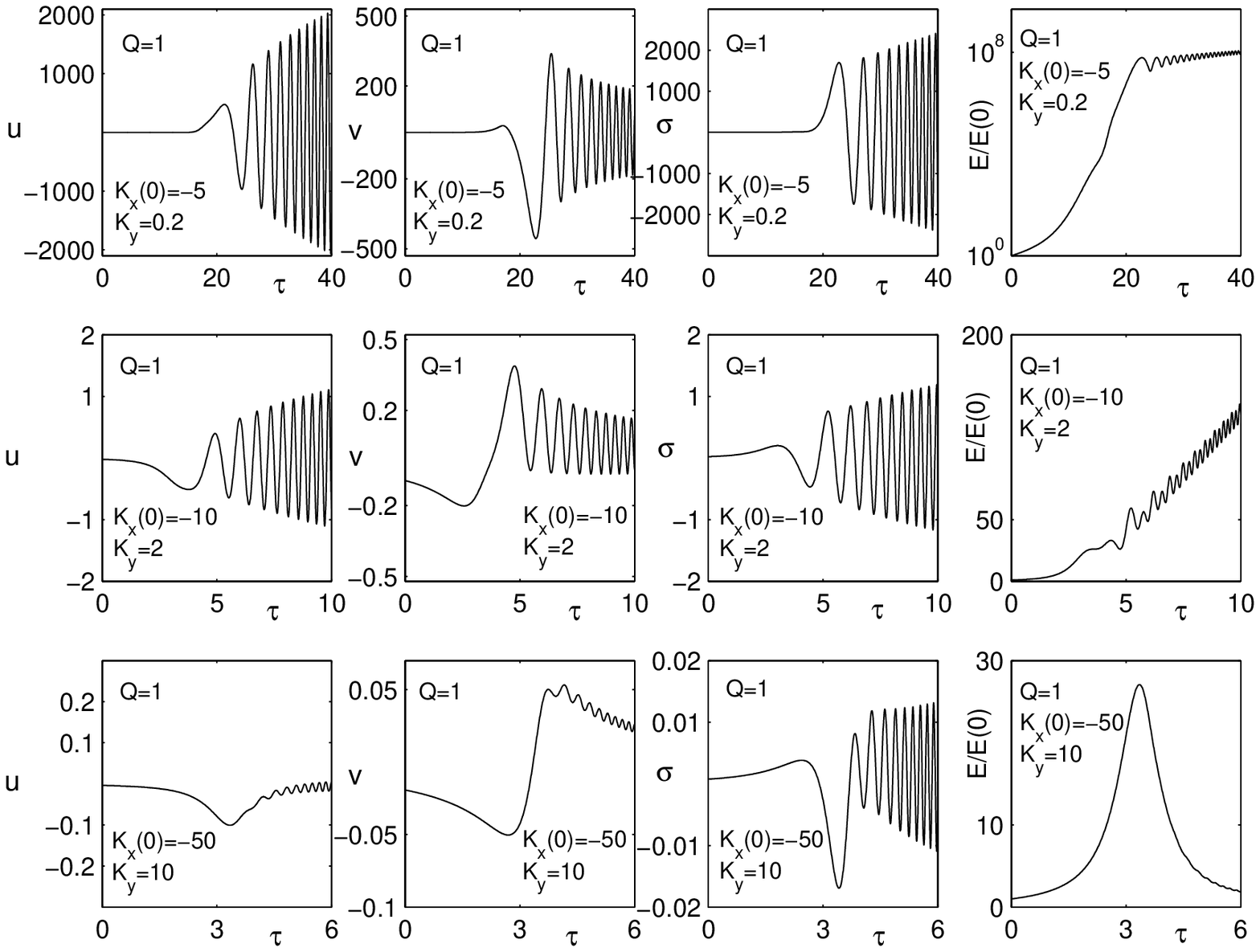}
\includegraphics[width=0.95\textwidth, height=0.4\textwidth]{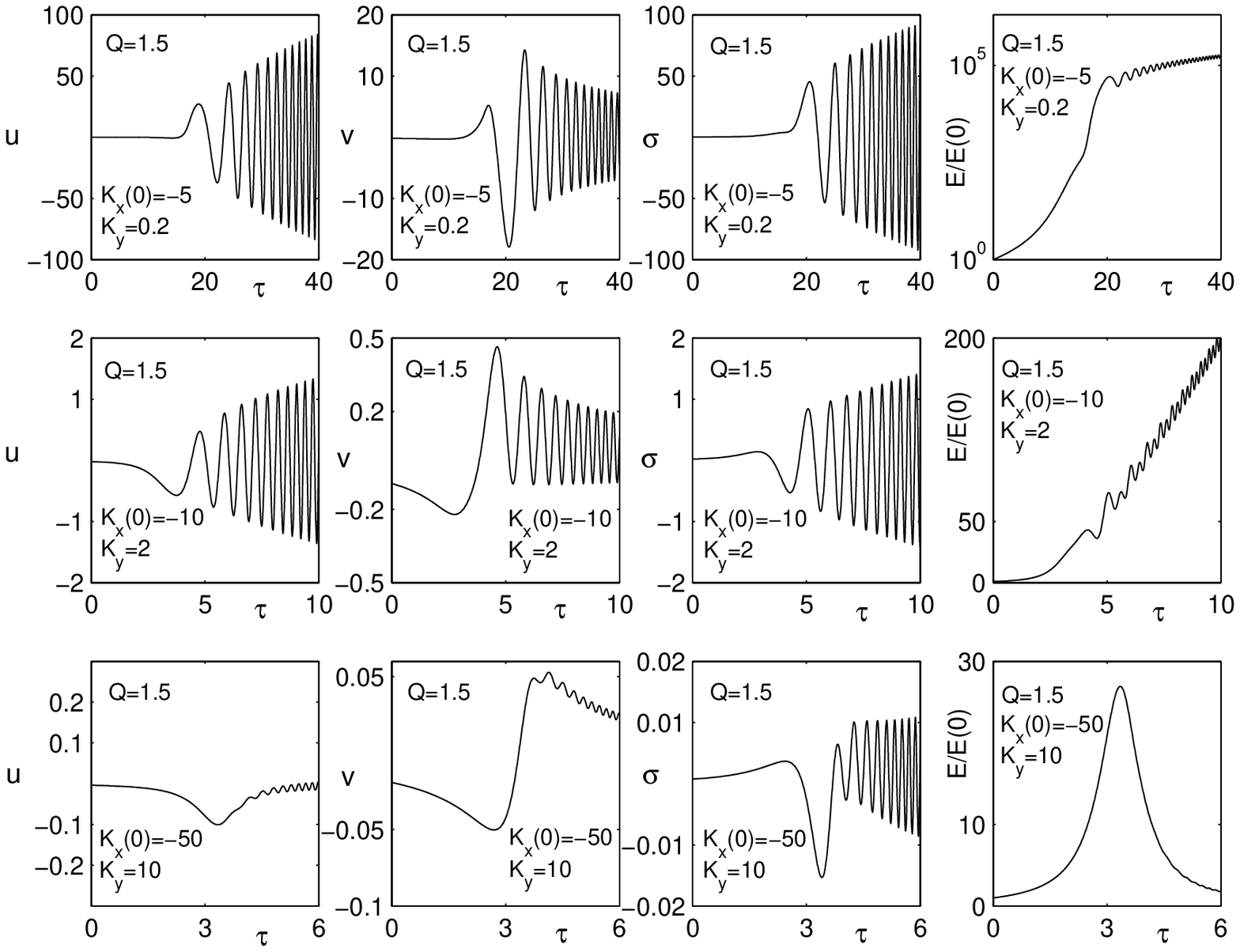}
\includegraphics[width=0.95\textwidth, height=0.4\textwidth]{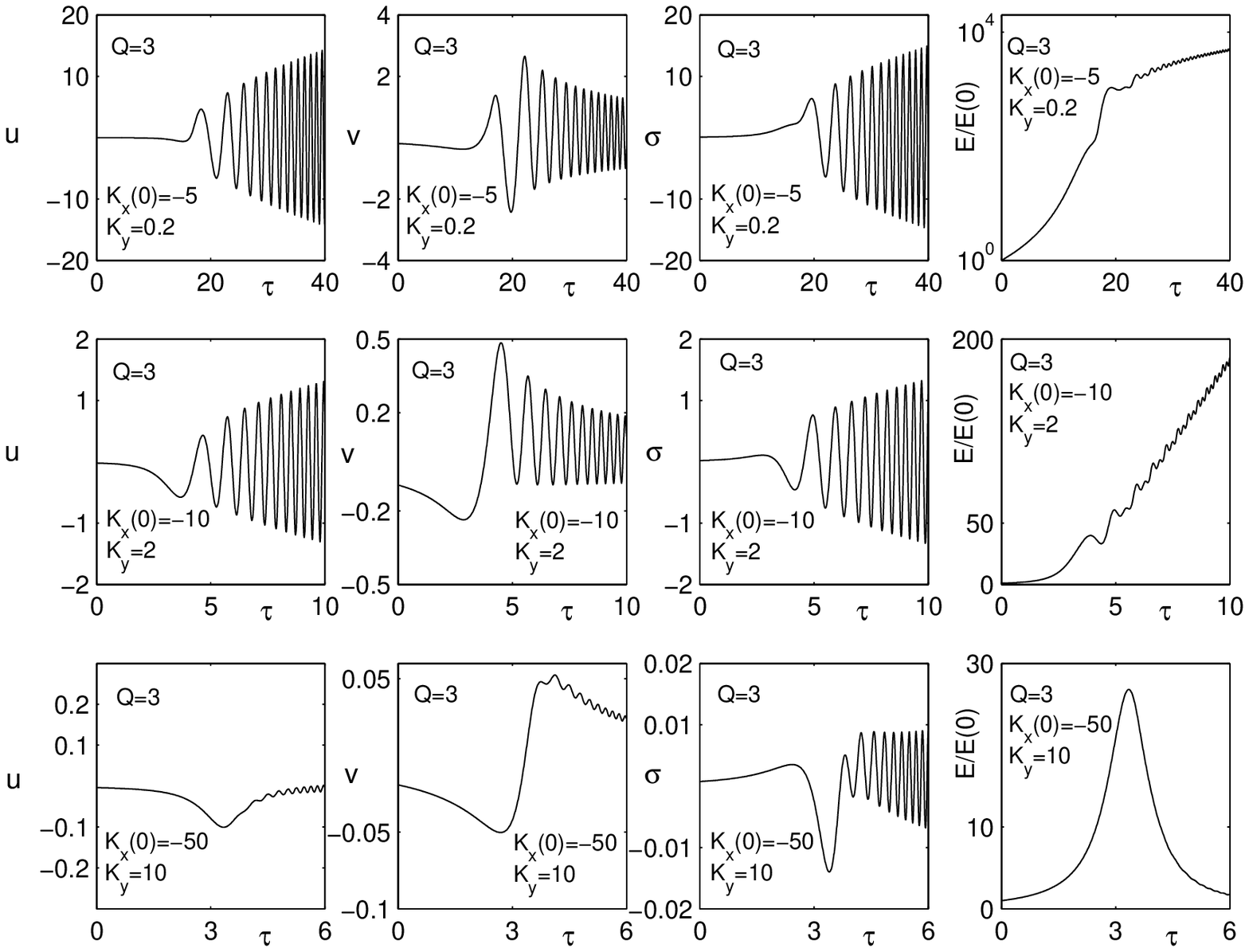}
\end{minipage}
\caption{Evolution of perturbed quantities pertaining to an
initially imposed leading pure vortex mode SFH at different $Q$,
$K_{x}(0)$ and $K_y$ (corresponding values are shown in each panel).
Before reaching the first unstable domain, the vortex mode SFH gains
energy from the mean flow and amplifies, but retains its aperiodic
nature. In the unstable domains, oscillations (SD waves) begin to
appear. The energy increases monotonically and then, after going
through the unstable domains, asymptotically linearly corresponding
to the generated SD waves (for $K_{y}Q<2$, the energy curves are in
logarithmic scale, so the flat part is actually linear growth). The
contribution of the vortex mode energy to the total perturbation
energy is negligible at this asymptotic stage. The vortex mode
simply dies down giving way to trailing SD waves. The wave
generation is very effective for $K_{y}Q<2$, moderate for
$K_{y}Q\sim 2$ and vanishing for $K_{y}Q \gg 2$.}
\end{figure*}

In fact, the (modified) initial value problem is solved by equations
(6-9) (or, equivalently, by equations (12-14)). The character of the
dynamics depends on the mode of perturbation inserted initially into
equations (6-9): pure SD wave mode (without admixes of aperiodic
vortices) or pure aperiodic vortex mode (without admixes of SD
waves).

Classification of perturbation modes that is widespread divides them
into vortical and divergent types. Such a classification coincides
with the described above physical classification in the case of
nonvortical (rigid) mean rotation, when the wave mode has zero
potential vorticity, but nonzero divergence and the vortex mode has
zero divergence, but nonzero potential vorticity. In the considered
quasi-Keplerian (i.e. strongly sheared) flow, the situation is
fundamentally different: the vortex mode may acquire divergent
nature and initially predominantly divergent wave mode may acquire
curl in the course of evolution.

In Fig.~2 we present the time-development of the parameter:
\[
~~~~~~~~~~~~~~~~~~~~~~~~\eta=\left| \frac{K_x(\tau) {\hat v}(\tau) -
K_y {\hat u}(\tau)}{K_x(\tau) {\hat u}(\tau) + K_y {\hat
v}(\tau)}\right|,
\]
which represents the ratio of the $z$-component of curl to
divergence, and its inverse value $1/\eta$ for initially imposed SD
wave and vortex mode SFHs respectively. In the case where initially
a predominantly divergent leading pure SD wave mode SFH with
positive frequency is inserted into equations (6-9) (the procedure
for selecting this type of initial condition is described in detail
in Nakagawa \& Sekiya 1992), it acquires curl at about the time of
entering the unstable (nonadiabatic) domain in Figs.~1,4, as seen in
Fig.~2(a). Fig.~2(b) shows that an initially inserted leading pure
vortex mode SFH acquires divergent nature at about the same time.

Thus, divergent (or vortical) perturbations in the quasi-Keplerian
flow (or, in a shear flow in general) represent some mix of vortex
and wave modes and classification of perturbations as vortical and
divergent may be misleading. So, we prefer the classification of
perturbations into wave and vortex modes and investigate dynamical
processes in terms of dynamics of these two modes.

\section{Transient growth and coupling of vortex and SD wave mode SFH{\sevensize s} -- numerical analysis}

We begin the numerical integration of equations (6-9), choosing as
an initial condition leading ($K_x(0)/K_y<0$) pure vortex mode SFH
without any admixes of SD wave mode SFHs. Such a selection of the
vortex mode is possible only far from the unstable domains, where
$|K_x(\tau)/K_y|\gg 1$ and the adiabatic condition $|d{\hat
\omega}(\tau)/d\tau|\ll {\hat \omega}^{2}(\tau)$ is met. The
procedure for selecting is described in detail in the Appendix of
Bodo et al. 2005. In Fig.~3, we present the subsequent evolution of
${\hat u}$, ${\hat v}$, ${\hat \sigma}$ and ${\hat E}/{\hat E}(0)$
for this kind of initial condition at different values of $Q$ and
$K_y$ (in these and other figures below, hats are omitted). The
sketch of the SFH's evolution/drift in wavenumber plane is given in
Fig.~4. We single out a leading SFH, for which $K_y < 2Q^{-1}$ and
that is located initially at point 1 far from the unstable domains
and coincides here, as mentioned above, with a pure vortex mode SFH
(henceforth, we take the azimuthal wavenumber $K_{y}$ to be positive
without loss of generality). As seen in this figure, this SFH drifts
along the $K_x$-axis in the direction denoted by the arrows ($1 \to
2 \to 3 \to 4 \to 5 \to 6$). The drift velocity ($= 2K_y |\hat{A}|$)
depends linearly on $K_y$. Initially, being in the adiabatic region,
the SFH gains energy from the mean flow due to the nonnormality and
amplifies algebraically, but retains its aperiodic nature. Then, the
dynamics becomes nonadiabatic -- the SFH reaches the unstable domain
where $\hat{\omega}^{2}(\tau)<0$ (point 2). From this point, a
temporal exponential growth and simultaneous excitation of the
corresponding SFH of SD wave mode take place -- at this stage of the
evolution, the linear coupling of vortex and wave mode SFHs is at
work (this phenomenon was found and thoroughly described for the
simplest shear flow in Chagelishvili et al. 1997 and for
non-self-gravitating Keplerian discs in Bodo et al. 2005). Then, the
vortex and the generated SD wave mode SFHs reach the intermediate
stable region (point 3) where $\hat{\omega}^{2}(\tau)>0$, pass it
and get again into the domain where $\hat{\omega}^{2}(\tau)<0$
(point 4). Further exponential growth of both vortex and SD wave
mode SFHs and, in addition, excitation of another SD wave mode SFH
by the vortex mode one occur until they cross this second unstable
domain (point 5). After that, the linear dynamics of the vortex and
SD wave mode SFHs become decoupled and adiabatic: the energy of the
vortex mode SFH dies down and the energy of the wave mode SFHs
increases. No further SD wave excitation is observed beyond point 5.

Here we have described the SD wave generation for $K_y<2Q^{-1},$
although it similarly occurs for $K_{y}\sim 2Q^{-1}$ (see Fig.~3),
except that the transient amplification of an initially imposed
vortex mode SFH is mainly due to the nonnormality, since the
unstable domains do not extend to such $K_y$. As a consequence, the
amplification amount and the amplitudes of generated SD wave mode
SFHs are several orders of magnitude lower than those for
$K_y<2Q^{-1}.$
\begin{figure}
\includegraphics[width=\columnwidth]{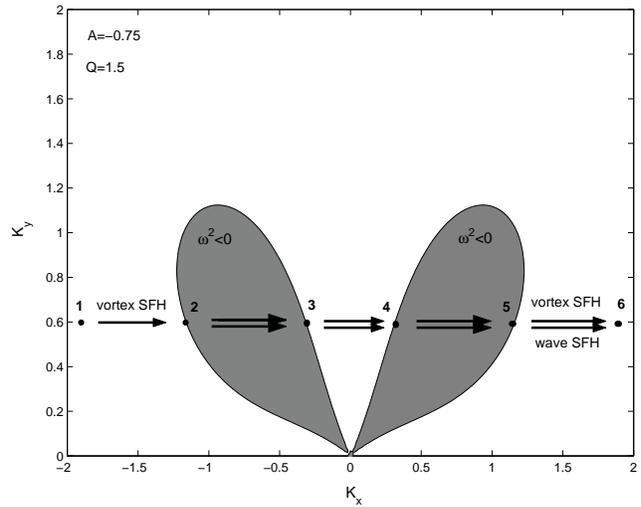}
\caption{Sketch of SFH's evolution in {\bf K}-plane. A leading
vortex mode SFH, located initially at point 1, drifts with time
along the $K_x$-axis in the direction denoted by the arrows $1 \to 2
\to 3 \to 4 \to 5 \to 6$. After crossing point 2, a SD wave mode SFH
appears. The first stage of the wave excitation takes place from
point 2 to point 3, the second one -- from point 4 to point 5.}
\end{figure}
\begin{figure}
\includegraphics[width=\columnwidth]{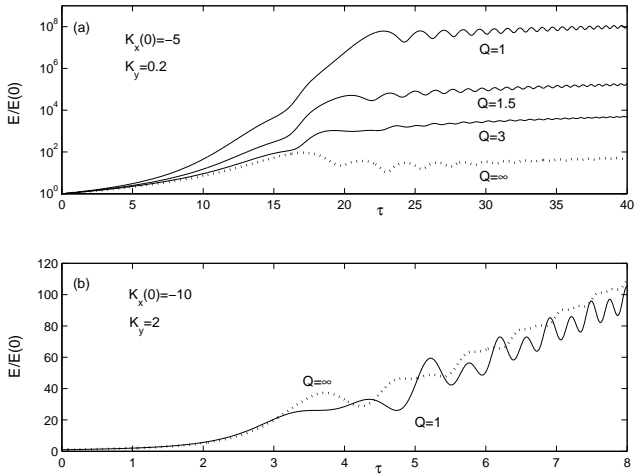}
\caption{Influence of self-gravity on the transient (swing)
amplification of a vortex mode SFH (dotted lines correspond to the
non-self-gravitating case). This influence is, as expected, largest
for $K_{y}Q<2$ (panel (a), all four curves are characterized by the
same $K_{x}(0)=-5,~K_{y}=0.2$ and identical initial values of
perturbed quantities) and negligible for $K_{y}Q\sim 2$ (panel (b),
$K_{x}(0)=-10,~K_{y}=2$).}
\end{figure}

Thus, the linear dynamics of a vortex mode SFH is followed by the
generation of the corresponding SD wave mode SFHs. This generated
SFHs eventually acquire a trailing orientation, since
$K_x(\tau)/K_y>0$ after leaving the nonadiabatic region (that
stretches roughly from point 2 to point 5 in Fig.~4). In the
nonadiabatic region, the characteristic timescales of the vortex and
SD wave mode SFHs are comparable and the perturbation modes cannot
be separated/distinguished. But with moving away from the
nonadiabatic region, modes get cleanly separated: the timescale of
the SD wave mode SFHs becomes much shorter than that of the vortex
mode SFH (the frequency of waves increases with time). One can
formally divide the energy evolution into two stages: the first
stage represents the transient amplification (both due to the
nonnormality and to the unstable domains) of the originally imposed
pure vortex mode SFH and excitation (and also subsequent exponential
amplification) of the corresponding SFHs of SD wave mode, and the
second one represents the algebraic growth of the generated SD wave
mode SFHs. The latter exhibit linear amplification at asymptotically
large times. In the absence of the wave excitation (e.g. for $K_y
\gg 2Q^{-1}$), this second stage describes decreasing energy of a
vortex mode SFH. Thus, newly created trailing SD wave mode SFHs in
the linear regime extract energy from the mean quasi-Keplerian flow
in contrast to a trailing vortex mode SFH that after leaving the
nonadiabatic region, gradually returns all its energy to the mean
flow. One can say that vortex mode perturbations act as a mediator
between the mean flow and waves. The energy needed for the wave
excitation is extracted from the shear and self-gravity with the
help of the vortex mode.

The following two subsections are devoted to the behaviour of vortex
mode perturbations for various values of Toomre's parameter $Q$.

\subsection{Effect of self-gravity on the transient growth of
vortex mode perturbations}

\begin{figure*}
\includegraphics[width=\textwidth, height=0.6\textwidth]{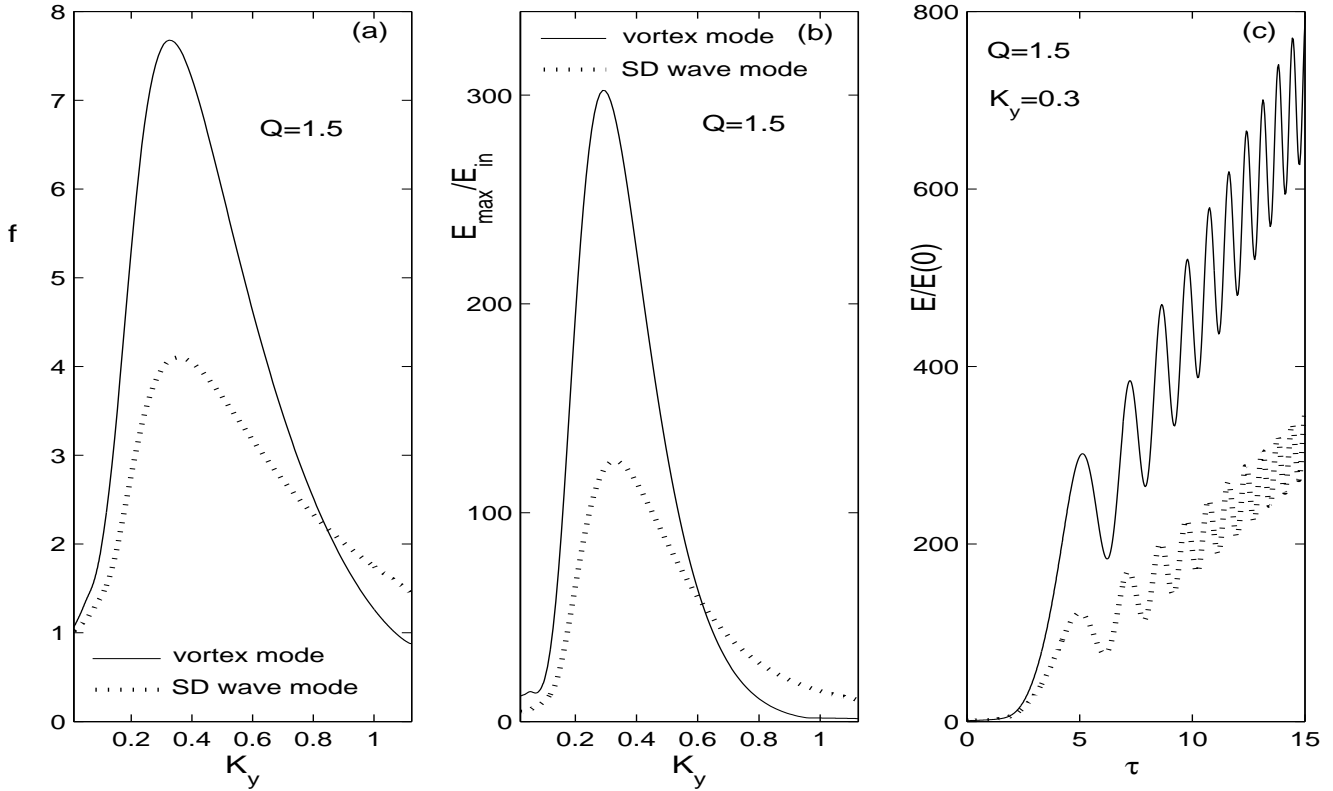}
\caption{Growth factors of density (panel (a)) and energy (panel
(b)) vs $K_y$. Also shown is the evolution of ${\hat E}/{\hat E}(0)$
for two cases (panel (c)). Solid curves in the panels correspond to
the initially imposed vortex mode SFH, dotted curves -- to the
initially imposed SD wave mode SFH. In all three panels, $Q=1.5$. In
panel (c), $K_y=0.3$. $\tau=0$ corresponds to point 2 in Fig.~4.}
\end{figure*}

In Fig.~5, we compare the time-development of ${\hat E}/{\hat E}(0)$
for initially imposed vortex mode SFHs in the presence and absence
of self-gravity for different values of $Q$ and $K_y$. It is clear
that the growth of vortex mode SFHs continues longer time and may be
several orders of magnitude stronger than in the
non-self-gravitating case ($Q \rightarrow \infty$, dotted lines in
the panels). In this case, the growth of vortex mode SFHs occurs
just at the leading stage ($K_x(\tau)/K_y<0$), on becoming trailing
($K_x(\tau)/K_y>0$) SFHs give back energy to the mean flow and
weaken \cite{Ch03}. In the self-gravitating case instead, the
amplification of SFHs continues into the trailing stage as well due
to the existence of the unstable domain at $K_x(\tau)/K_y>0$ (see
Figs.~1,4). From Fig.~5, one can see that at $Q=1$ and $K_y=0.2$, a
vortex mode SFH grows about $10^6$ times stronger than in the
non-self-gravitating case; at $Q=1.5$ and $K_y=0.2$ -- about $10^4$
times stronger; at $Q=3$ and $K_y=0.2$ -- about $10^2$ times
stronger;  at $Q=1$ and $K_y=2$, i.e., at $K_{y}Q \sim 2$, the
growth is the same as in the non-self-gravitating case. In any case,
one can conclude that self-gravity provides a substantial
enhancement of the transient growth of vortex mode perturbations,
thereby making them active participants in dynamical processes.
This, in turn, shows that the bypass mechanism may play a part in
the onset of turbulence in self-gravitating discs.

In order to better understand the role of the vortex mode, it is
interesting to carry out a comparative analysis of the transient
(swing) amplification of the vortex and SD wave modes. First we
define a density growth factor $f$ for SFH initially located at
point 1, as a ratio of the absolute values of $\hat {\sigma}(\tau)$
after (at point 5) and before (at point 2) crossing the unstable
domains in Fig.~4, $f\equiv|\hat {\sigma}(\tau'')|/|\hat
{\sigma}(\tau')|$, where $\tau''$ and $\tau'$ are the moments
corresponding to points 5 and 2 respectively. A similar growth
factor for coherent wavelet solutions was used by Kim \& Ostriker
(2001), but they took its logarithm. In Fig.~6(a), we present this
parameter computed separately for the initially imposed vortex and
SD wave mode SFHs as a function of the dimensionless azimuthal
wavenumber $K_y$ at $Q=1.5$. The initially imposed SD wave mode SFH
has a certain (positive) sign of frequency. As seen in this panel,
in the dynamically important regions (i.e., for such values of
$K_y$, at which both modes experience maximum transient growth), the
growth factor for the vortex mode SFH is almost two times larger
than that for the SD wave mode one. An analogous comparison is made
in Fig.~6(b). Here we display the ratio of the maximum value
achieved by the energy $\hat {E}(\tau)$ during transient
amplification in the unstable domains to its initial value on
entering these domains (i.e., at point 2) computed separately for
the imposed at point 2 vortex and SD wave mode SFHs as a function of
$K_y$, similar to what is done in Fig.~6(a). But now, as distinct
from the first case, for the wave mode SFH we choose initial
conditions at point 2 in such a way as to obtain the largest
possible amplification of the wave energy in the transient growth
(swing) phase for its fixed initial value, i.e., we take transiently
most unstable wave mode SFH.  The situation is similar to the above
one: the energy amplification factor for the vortex mode SFH is more
than two times greater than the largest possible energy
amplification factor for the SD wave mode SFH. In Fig.~6(c), we
present the parallel evolution of the energies of the initially (at
point 2) imposed vortex and maximally amplified wave mode SFHs for
$K_y=0.3$, at which the energy growth factors of both modes during
swing phase are the largest (see Fig.~6(b)). Both SFHs start with
the same energy. This panel shows that the energy corresponding to
the initially imposed vortex mode SFH remains about two times larger
than that corresponding to the SD wave mode SFH at all times.

From Fig.~6, it is evident that the vortex mode prevails over the SD
wave mode in two respects: in the transient amplification and wave
generation (by the wave generation for the initially imposed SD wave
mode SFH we mean wave amplification due to the over-reflection
mechanism; see Nakagawa \& Sekiya 1992 for details). The latter
follows from the asymptotic stage at large times, when both energy
curves become linear (see Figs.~3,6(c)) with inclinations
proportional to the square of the amplitudes of generated SD wave
mode SFHs after crossing the unstable domains. We see that the
energy of SD wave mode SFHs generated by the initially imposed
vortex mode one at this asymptotic stage is about two times larger
than that of the generated by the initially imposed SD wave mode
SFH.

\subsection{Ways of SD wave generation}

There are two ways of SD wave generation in the considered here disc
flow. The first is a direct and well-known over-reflection
mechanism: inserting a leading SD wave mode SFH in equations (6-9),
one can get the energy dynamics represented by the dotted curve in
Fig.~6c. The curve describes the energy growth in the unstable
domains that is followed by a linear growth of the total energy of
the resulting over-reflected and (over)-transmitted trailing SD wave
mode SFHs at large times.

Another way of the generation of trailing SD wave mode SFHs is by
means of vortex mode SFHs: leading pure vortex mode SFHs can
effectively excite trailing SD wave mode ones due to the mode
coupling phenomenon. Figure~6 shows that this second way of SD wave
generation is about two times more effective than the first one.

We go on to calculate the amplitudes of SD waves generated due to
mode coupling (more precisely, the amplitudes for the gravitational
potential perturbations. The amplitudes for other quantities can
afterwards be found easily). Insert an adiabatic leading vortex mode
SFH into equation (14). Then after passing the nonadiabatic region
(in the other adiabatic region at $\tau \rightarrow \infty$) this
solution goes over to the superposition of a trailing SFH of the
vortex mode and generated SD wave mode SFHs:
\[
{\hat \phi}(\tau)={\hat \phi}^{\rm v}(\tau)+{\hat \phi}^{\rm
w}(\tau)= -\frac{4}{QK(\tau){\hat
\omega}^2(\tau)}\left(\hat{\Omega}_{0}+\frac{2{\hat
A}K_y^2}{K^2(\tau)}\right){\hat {\cal I}}+
\]
\[
~~~~~~~~~+\frac{a}{Q\sqrt{{\hat \omega}(\tau)}}e^{
-i\int^{\tau}{\hat
\omega}(\tau')d\tau'}+\frac{a^{\ast}}{Q\sqrt{{\hat \omega}(\tau)}}
e^{i\int^{\tau}{\hat \omega}(\tau')d\tau'},
\]
where $a$ and $a^{\ast}$ are the amplitudes of generated SD wave
mode SFHs. The latter come in complex conjugate pairs with different
signs of frequencies and, hence, propagating in the opposite
directions. In Fig.~7, we plot the numerically obtained $|a|$ as a
function of $K_y$ at $Q=1.5$ and in Fig.~8 the same for different
values of $Q$. In both figures $\hat{\cal I}$ is set to unity. The
procedure for the calculations is analogous to that employed by
Nakagawa \& Sekiya (1992) to study the over-reflection of SD waves.
\begin{figure}
\includegraphics[width=\columnwidth]{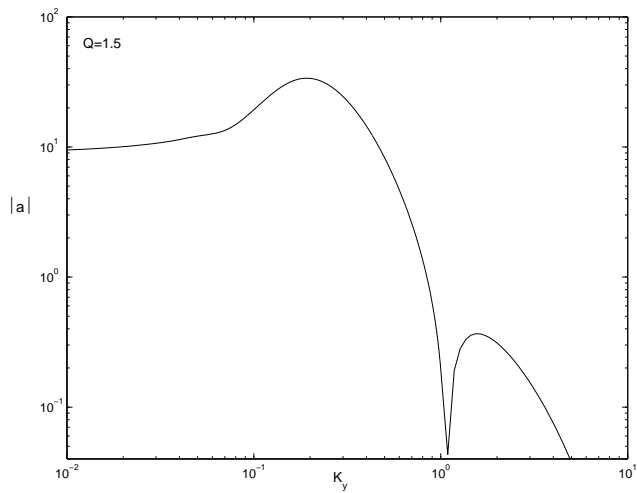}
\caption{The amplitude $|a|$ of trailing SD wave mode SFH generated
by a leading vortex mode SFH vs $K_y$ at $Q=1.5$. }
\end{figure}
\begin{figure}
\includegraphics[width=\columnwidth]{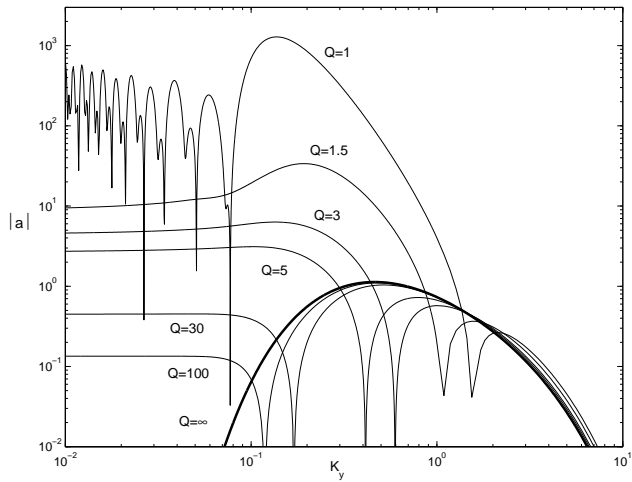}
\caption{Same as in Fig.~7, but for different values of $Q$,
including the non-self-gravitating case ($Q \rightarrow \infty $).}
\end{figure}

Let us analyse the curves in Figs.~7,8. The maximum value of $|a|$
is achieved for $K_{y}\sim O(0.1)$, as at such $K_y$ a SFH drifts
slowly in \textbf{K}-plane, consequently, slowly crosses the
unstable domains and has more time for transient growth. The
cavities in these figures are due to the crossing of two wave
excitation domains by the SFH (see Fig.~4) and, therefore, due to
the existence of two, more or less independent, stages of the wave
excitation/generation: the resulting (after leaving both unstable
domains) wave mode SFH is a superposition of SFHs generated at these
stages. At $Q=1.5$, this interference is destructive close to
$K_{y}=1$ and results in the cavity (Fig.~7). As one can see from
Fig.~8, the cavity point occurs at different $K_y$ for different
$Q$. At small values of $Q$, the number of cavity points increases
(see curve for $Q=1$ in Fig.~8), as destructive interference happens
at different $K_y$. In the non-self-gravitating case ($Q \rightarrow
\infty$), there is no cavity point, as in this case there is just
one stage of the wave excitation and, therefore, the interference
phenomenon is absent.

\section{Summary and Discussions}

Studying flow nonnormality induced transient phenomena in thin
self-gravitating astrophysical discs, we have concentrated on the
dynamics of vortex mode perturbations. The linear dynamics of
perturbations has been investigated by means of so-called nonmodal
approach, which consists in tracing the dynamics of SFHs (see
equation (5)). SFHs represent the simplest/basic ``elements'' of the
dynamical processes at constant shear rate and greatly help to grasp
transient growth and coupling of perturbation modes. It has been
shown that self-gravity considerably alters the dynamics of vortex
mode SFHs -- their transient growth may be several orders of
magnitude stronger than in the non-self-gravitating case and 2-3
times larger than the transient growth of the wave mode (see
Figs.~5,6).

The evolution of vortical and wave type perturbations has recently
been studied by Wada, Meurer \& Norman (2002) in high-resolution
numerical simulations of two-dimensional hydrodynamic turbulence in
certain galactic disc flows. These simulations clearly demonstrate
that vortical/solenoidal perturbations are equally important
together with spiral density wave/compressible perturbations in
determining the properties (spectra) of the resulting
gravitoturbulent state. They also suggest, based on their simulation
results, that self-sustained turbulence can also occur in the case
of self-gravitating Keplerian rotation. Their work supports the
conclusion that the described here gravity-enhanced transient growth
of a vortex mode perturbation is a key factor in the simulated
self-sustained turbulence and, hence, the vortex mode perturbation
itself -- a key participant. Consequently, self-gravity, or
gravitational instability, might allow for the onset of
two-dimensional hydrodynamic turbulence in astrophysical disc flows
and the bypass mechanism of the onset of turbulence (elaborated by
the hydrodynamic community in the 90s of the last century), may play
a part in this process.

Another relevant to the present paper work is that of Gammie (2001).
In this paper, turbulence and angular momentum transport in
self-gravitating discs are studied numerically in the shearing sheet
approximation. One has to note that the initial white noise
distribution adopted by Gammie (2001), in fact, includes also vortex
mode perturbations, as the potential vorticity of the initially
imposed white noise is nonzero, i.e., the initial perturbation is a
mixture of vortex and wave modes. However, as mentioned in the
Introduction, the identification/separation of the modes and the
separate study of their properties have been left out of analysis.
In the case of such a mixture, vortex mode perturbations grow
transiently and at the same time generate zero vorticity
perturbations (i.e., SD waves) via the described here linear
mechanism. The resulting turbulence can be more "violent" than the
zero vorticity one, i.e., turbulence in which the basic elements are
SD waves, and therefore the angular momentum transport can be more
intense because of the larger transient growth factors of vortex
mode perturbations. There should also be differences in the
statistical properties (energy spectra) of these two kinds of
turbulence. (This question is addressed, in part, in Wada, Meurer \&
Norman 2002. However, they make somewhat different from ours
classification of perturbation modes). Gammie's analysis actually
concentrates only on the question of locality of angular momentum
transport in a gravitoturbulent state and not on the investigation
of the relative contributions/fractions of vortex and SD wave mode
perturbations in shear stresses governing angular momentum transport
in discs.

The described linear coupling of vortex and wave modes, which is
caused by the differential character of the disc flow, is
asymmetric: vortex mode perturbations are able to generate wave mode
ones, but not vice versa. The considered system conserves potential
vorticity and it is obvious that vortex mode perturbations, having
nonzero vorticity, are able to excite SD wave mode perturbations
having zero potential vorticity. This asymmetric coupling lends
additional significance to the vortex mode as a participant in SD
waves and shocks manifestations in astrophysical discs.

\section*{Acknowledgments}
This work is supported by the International Science and Technology
Center (ISTC) grant G-1217. We would like to thank the anonymous
referee for helpful comments.

\end{document}